\documentclass[twocolumn,prl,amsmath,showpacs]{revtex4}
\usepackage{graphicx}
\usepackage{dcolumn}%
\usepackage{bm}%
\usepackage{amssymb}
\usepackage{amsmath}
\usepackage{color}

\def\be{\begin{equation}}
\def\ee{\end{equation}}

\begin{document}

\title{Disorder-assisted melting and the glass transition in amorphous solids }
\author{Alessio Zaccone and Eugene M. Terentjev}
\affiliation{Cavendish Laboratory, University of Cambridge, JJ Thomson Avenue,
Cambridge CB3 0HE, U.K.}
\date{\today}

\begin{abstract}
\noindent The mechanical response of solids depends on temperature because the
way atoms and molecules respond collectively to deformation is affected at
various levels by thermal motion. This is a fundamental problem of solid state
science and plays a crucial role
in metallurgy, aerospace engineering, energy. In glasses the vanishing of
rigidity upon increasing temperature is the reverse process of the glass
transition. It remains
poorly understood due to the disorder leading to nontrivial (nonaffine)
components in the atomic
displacements. Our theory explains the basic mechanism of the melting
transition of
amorphous (disordered) solids in terms of the lattice energy lost to this
nonaffine motion,
compared to which thermal vibrations turn out to play only a negligible role.
It predicts the square-root vanishing of the shear modulus $G\sim\sqrt{T_{c}-T}$ at criticality observed in the most recent numerical simulation study.
The theory is also in good agreement with classic data on melting of amorphous
polymers
(for which no alternative theory can be found in the literature) and offers new
opportunities in materials science.
\end{abstract}

\maketitle

The phenomenon of the transition of a supercooled liquid into an amorphous
solid
has been studied extensively and many theories have been proposed in the past, starting with the Gibbs-DiMarzio theory~\cite{gibbs}.
All these theories focus on the fluid to solid aspect, coming to
this glass transition from the liquid side (supercooling). However, it is only
one facet of the problem. For the reverse process, i.e. the melting of the
amorphous solid into a liquid, no established theories are available.
%although it plays
%an equally important role in our understanding of the underlying physics and is
%equally important in practice.

\begin{figure}
\includegraphics[width=.75\linewidth]{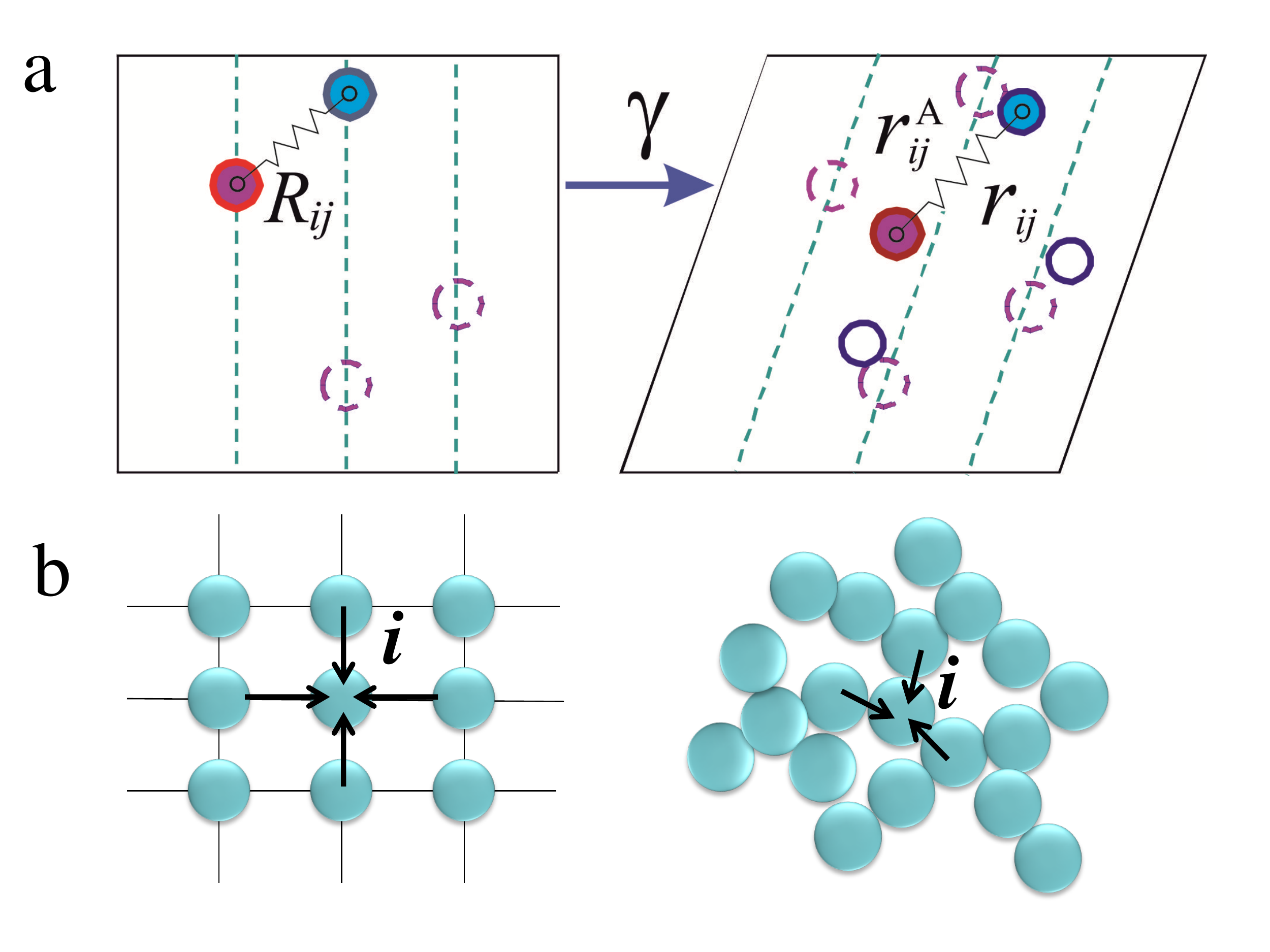}
\caption{\label{figure1} \textbf{a},
If the deformation was affine, the particles labeled with darker color would
still lie on the dashed lines also in the deformed (right) frame. Because
of nonaffinity, they do not. \ \textbf{b}, In an ordered crystal lattice (left)
the forces (arrows) transmitted to any particle $i$ by its nearest neighbors
upon deformation cancel each other and the net force acting on $i$ is zero
by symmetry. In a disordered lattice (right) the forces
transmitted upon deformation do not balance, and a net
force acts on $i$ which has to be balanced by a
nonaffine motion, to further lower the lattice potential energy in order
to preserve the mechanical equilibrium.}
\end{figure}

The problem of describing the melting transition into a fluid
state~\cite{born,edwards,chaikin,science,frenkeld,rastogi,brillouin,frenkel}
is complicated in amorphous solids by the difficulties inherent in describing
the elasticity down to the atomistic
level (where the thermal fluctuations take place).
It is
well known that the standard (Born-Huang) lattice-dynamic theory of elastic
constants, and also its later developments~\cite{parodi}, breaks down on the microscopic scale. The reason is that its basic
assumption, that the macroscopic deformation is
\emph{affine} and thus can be down-scaled to the atomistic level, does not
hold~\cite{alexander}. Atomic displacements in amorphous solids are in fact
strongly \emph{nonaffine}~\cite{alexander,lubensky,barrat,mackintosh}, a
phenomenon illustrated in
Fig.~\ref{figure1}.
Nonaffinity is caused by the lattice disorder: the forces transmitted
to every atom by its bonded
neighbors upon deformation do not balance, and the resulting non-zero force can
only be equilibrated by an additional nonaffine displacement, which adds
to the affine motion dictated by the macroscopic strain.

Recently, it has been shown~\cite{mezard} that nonaffinity could play a role in
the melting
of model amorphous solids, although the basic interplay between nonaffinity,
thermal expansion, and thermal vibrations remains unclear. With a number of
other models of the glass
transition, such as mode-coupling theories, there is also an
issue when they rely on liquid-state theory for relating the stress
tensor to local fluctuations in the solid state. These
theories predict that the shear modulus $G$ of \emph{athermal}
hard-sphere colloids
remains finite at the glass transition~\cite{siebenbuerger} and that it jumps
discontinuously to zero upon decreasing the packing fraction $\phi$. This
scenario does not agree, however, with \emph{thermal} systems, where both
simulations~\cite{barrat88} and experiments~\cite{klix} show that the
vanishing of $G$ is continuous with $T$. The glass transition temperature
$T_{c}$ could therefore be estimated for thermal systems using the Born melting
criterion~\cite{born}, i.e. setting $G(T)=0$~\cite{thorpe1} and solving for
$T_{c}$, if a theory for the low-frequency $G$ in the solid state
is available.

Here we discover
a much more basic effect that nonaffinity of deformation causes
in the melting, where marginal stability controls the response through the
bond-connectivity, in turn affected by both $T$ (via thermal expansion) and
disorder. The resulting
framework allows us to link lattice disorder, connectivity and
thermal vibrations into the first description of the classical
experiment~\cite{schmieder} on $T$-dependent low-frequency shear modulus of amorphous polymers.

Let us start from the basic assumptions of Born-Huang lattice
dynamics~\cite{born}. The free
energy density of affine deformation is given by the following harmonic lattice
sum:
$F_{A}=\frac{1}{2V}\sum_{ij}\left(\partial^2 U/\partial
\underline{r}_{ij}^{2}\right)_{R_0} (\underline{u}_{ij}^{A})^{2}$, which runs
over all bonded atom pairs $ij$. Here $U(\underline{r}_{ij})$ is the
pair-interaction potential and the
vector
$\underline{u}_{ij}^{A}=\underline{r}_{ij}^{A}-\underline{R}_{ij}=\underline{\underline{\eta}}\cdot\underline{R}_{ij}$
denotes the affine displacement, with
$\underline{\underline{\eta}}$  the macroscopic strain tensor and
$\underline{R}_{ij}$ the bond vector in the undeformed frame.  $R_{0}$ is the
equilibrium lattice constant in the undeformed frame, at which one evaluates
the lattice spring constant $\kappa=\left(\partial^2 U/\partial
\underline{r}_{ij}^{2}\right)_{R_{0}}$. {Without loss of generality, we focus
on \emph{shear} strain $\eta_{xy}\equiv \gamma$.} The lattice sum can be
evaluated upon
introducing the average number of bonds per atom $z$, and in the affine
approximation~\cite{zaccone}: $F_{A} = \frac{2}{10\pi}
(\kappa/R_0) \phi z\gamma^{2}$. Here $\phi=v N/V$ is the
packing fraction occupied by the atoms or building blocks of the
solid.

In general, the contribution to $F$ due to
thermal phonons is given by: $F_{T}=-kT\ln\sum_{n}^{\infty} \exp {(-\hbar
\omega_{n}(n+\frac{1}{2})/kT)}$, where $n$ labels the eigenmodes.
If $kT\gg \hbar \omega_{\mathrm{max}}$, one has: $F_{T}=-(3N/V)kT\ln(kT/\hbar
\overline{\omega})$, where $\overline{\omega}$ is defined such that $\ln
\overline{\omega}$ is equal to the average value of $\ln \omega$. The
contribution of the thermal phonons can be
written as: $F_{T} \approx -(3N/V)kT\theta\gamma^{2}$ where the
non-dimensional factor
$\theta=-(\partial^{2}/\partial\gamma^{2})_{\gamma\rightarrow 0} \ln \hbar
\overline{\omega}/kT$ has been demonstrated~\cite{frenkel} to be of order unity
when the harmonic potential dominates the pair interaction potential. This
gives a good estimate: $F_{T} \approx -(3N/V)kT \gamma^{2}$.

\begin{figure}
\includegraphics[width=.75\linewidth]{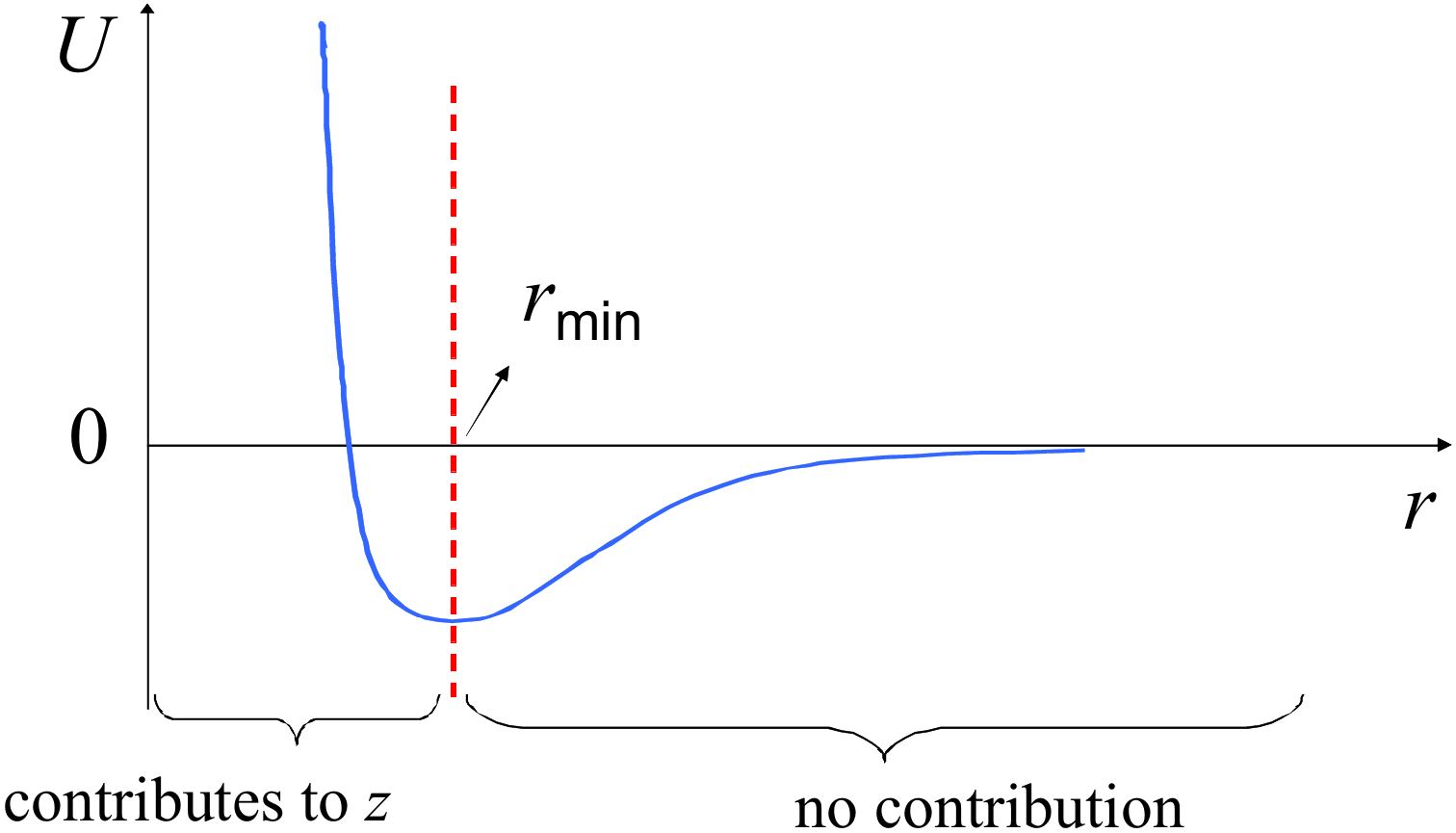}
\caption{\label{figure2} Schematic of the criterion used to define
the contribution of the weaker Lennard-Jones interactions to
the total number of mechanical bonds $z$. Only pairs of particles that lie
within the soft repulsive part of the potential (to the left of the minimum)
contribute to the $z$ counting.}
\end{figure}

The number of mechanical bonds per atom $z$ requires a careful definition in
amorphous thermal systems. If there were only covalent bonds, then $z$ is
obviously just equal to the number of covalent bonds per atom.
However, in addition to covalent bonds, weaker interactions are present, such as those between two monomers of different chains in
polymer systems. Such interactions are of van der Waals
nature and they can be modeled by the Lennard-Jones (LJ) potential. In
that case, it is important to distinguish quantitatively between these
contributions in the total $z$.
Here we propose the following criterion that allows us to calculate these terms
unambiguously.
With a glassy polymer in mind (although without any loss of generality), we
shall assume that
a contribution to $z$ arises from inter-chain interactions whenever the two
monomers are at a mutual separation $r\leqslant r_{\mathrm{min}}$, where
$r_{\mathrm{min}}$ is the minimum of the LJ potential well, see
Fig.~\ref{figure2}. Although this procedure might slightly underestimate the
actual absolute value of $z$, the underestimation has no effect on the
change $\delta z$ upon varying $T$ which is the quantity that matters in our
calculation. Later in the text, when we compare the model predictions with a
particular experiment on polymer glasses~\cite{schmieder}, we shall derive the
explicit values for the two components contributing to the total average
$z$: the
one due to intra-chain covalent bonds $z_{\rm co}$ and
the one due to inter-chain LJ bonds, which we denote $z_{\rm LJ}$.

{Having defined the total $z$,
we now need to relate it to $T$.
Upon introducing the thermal expansion coefficient,
$\alpha_{T}=\frac{1}{V}\left( \partial V / \partial T \right)$, and replacing
the volume $V$ via $\phi=v N/V$, after integration we obtain
$\ln(1/\phi)=\alpha_{T}T+const$ (later we shall need to estimate this constant,
obtaining $C \sim 0.48$). Now $z$ can be estimated as a function of $\phi$ by
introducing the radial distribution function (rdf) $g(r)$. Since the average
connectivity due to covalent bonds remains fixed, only the weaker contact bonds
contributing to $z_{\rm LJ}$ are changing upon increasing the packing fraction
$\phi$ by $\delta\phi$. The increment $\delta z$ can be
calculated in full analogy with soft-sphere systems where only the repulsive
part of the potential is active. This increment is given by~\cite{wyart}:
$\delta z \sim
\int_{1}^{1+\delta\phi}\xi^{2}g(\xi)d\xi$, where $\xi$ is the normalized
center-to-center distance. The rdf in the repulsive part is dominated by
$g(\xi)\sim\sqrt{\xi-1}$, as shown in theory~\cite{wyart} and
simulations~\cite{ohern}. To keep things analytical, here we neglect the
thermal broadening of the rdf
 (which could be calculated
only using involved replica techniques~\cite{zamponi}). This particular
simplifying approximation should still work for relatively low-$T_{c}$ polymer
glassy systems, as we are going to verify below, but
certainly has to be adjusted when dealing with inorganic glasses which have an
order of magnitude higher $T_{c}$.}  The increment $\delta z$ has to be
measured from the point
where the system is marginally stable, i.e. $z=z_{c}$ at $\phi=\phi_{c}$, and
from the integral we obtain: $z-z_{c}\sim\sqrt{\phi-\phi_{c}}$. In the affine
approximation, the solid becomes marginally stable only in the limit
$z_{c}\rightarrow0$ and $\phi_{c}\rightarrow0$, and hence we have
$z\sim\phi^{1/2}$. Using the earlier relation between $\phi$ and $\alpha_{T}$
we obtain: $z\sim e^{-\alpha_{T} T/2}$. Substituting $z$ and $\phi$ in
$F_{A} + F_T$ we now can write the full
expression for the shear modulus in the affine approximation,
$G_A=\partial^{2}(F_{A}+F_{T})/\partial\gamma^{2}$,  yielding:
\begin{equation}\label{eq:aff}
G_A(T) = \frac{2}{5\pi}\frac{1}{R_{0}^{3}} (\kappa R_{0}^{2}
e^{-(3/2)\alpha_{T}T}-kTe^{-\alpha_{T}T}).
\end{equation}
The Born criterion of melting~\cite{born}, is given by Eq.(\ref{eq:aff}) set to
zero:
$\kappa R_{0}^{2}  = kTe^{\alpha_{T}T/2}$. We shall see later that $\alpha_{T}T
\ll 1$ and, remarkably, this relation reproduces the Lindemann
criterion~\cite{lindemann}, which uses equipartition to state that melting
occurs
when the average vibrational energy of a bond equals $kT$. The Lindemann
criterion grossly overestimates melting
temperatures for amorphous
solids~\cite{alexander};  also, Eq.(\ref{eq:aff})
cannot capture the vanishing of rigidity as seen in the melting of glassy
polymers~\cite{schmieder}. It turns out that to describe the
melting of amorphous solids one has to account for nonaffine deformations
in the lattice dynamics.

The shear modulus accounting for nonaffine deformations is derived
in~\cite{lemaitre} as a lattice sum:
$G=
G_{A}-G_{NA}=G_{A}-\sum_{i}\underline{f}_i\sum_{j}\underline{\underline{H}}_{ij}^{-1}\underline{f}_j$,
where
$\underline{\underline{H}}_{ij}=(\partial^{2}
F/\partial\underline{r}_{i}^{2}\partial
\underline{r}_{j}^{2})_{\gamma\rightarrow0}$ is
the dynamical matrix of the solid. The vector
$\underline{f}_{i}$
measures the increment of force acting on an atom in response
to the deformation (here $\gamma\equiv\eta_{xy}$) of its environment. It can be
shown that for
harmonic pair potential~\cite{lemaitre}:
$\underline{f}_{i}=-R_{ij}\kappa\sum_{j}\underline{e}_{ij} e_{ij}^x e_{ij}^y$,
where
$\underline{e}_{ij} $ is the unit
vector along the bond connecting two atoms $i$ and $j$. Since the sum runs over
bonds to the nearest-neighbors $j$ of the atom $i$, it is evident that in a
perfect crystal for each bond involving $i$ there is a mirror-image bond across
a reflection plane of the crystalline lattice. Therefore, every bond in the sum
cancels with its mirror-image, and $\underline{f}_{i}=0, ~\forall i$ in
most crystal lattices. As a result, $\underline{f}_{i}$ is nonzero only with
lattice disorder, see Fig.~\ref{figure1}b.
The nonaffine correction to the elastic free energy
arises to ensure that mechanical equilibrium, which disorder tends to
compromise, is preserved upon deformation. It has been evaluated, in
mean-field approximation, for random assemblies of
harmonically bonded particles~\cite{enzo,jamie}:
\begin{equation} \label{eq:nonaff1}
G= G_{A}-G_{NA}=\frac{2}{5\pi}\frac{\kappa }{R_0}\, \phi (z-z_{c}).
\end{equation}
It should be noted that this expression has been derived for harmonic
pair potentials and is therefore valid
for any potentials with an attractive minimum.

The nonaffine contribution is encoded in Eq.~(\ref{eq:nonaff1}) in the term
proportional to $z_{c}$ which
expresses the internal energy
required to fuel the nonaffine motions necessary for the preservation of
mechanical equilibrium against the effect of disorder. With purely
central-force interactions in $d$ dimensions, the shear modulus vanishes at
$z_{c}=2d$ ($z_c=6$ in 3D) because the nonaffine term is proportional to the
number of degrees of freedom
that can be involved in the nonaffine energy relaxation. This is consistent
with the classical
Maxwell criterion for marginal stability with purely central
forces: $G\sim(z-6)$. In general, $z_{c}$ defines the critical coordination at
which the
lattice is no longer rigid because all the lattice potential energy is
``spent'' on sustaining the nonaffine motions and no energy is left to support
the elastic response to deformation. Using again: $\ln(1/\phi)=\alpha_{T}T+C$,
we arrive at: $\ln(\phi_{c}/\phi)=\alpha_{T}(T-T_{c})$. The corresponding
relation $z-z_{c}\sim\sqrt{\phi-\phi_{c}}$  can be manipulated into:
\begin{equation} \label{eq:wyart}
\ln(\phi_{c}/\phi)=-\ln[1+(z-z_{c})^{2}/\phi_{c}].
\end{equation}
Combining this with the  relation for $\phi (\alpha_T)$ we obtain
$\ln[1+(z-z_{c})^{2}/\phi_{c}]=\alpha_{T}(T-T_{c})$, and finally arrive at the
condition: $z-z_{c}=\sqrt{\phi_{c}[e^{\alpha_{T}(T_{c}-T)}-1]}$. Substituting
it in Eq.(\ref{eq:nonaff1}), we obtain:
\begin{equation}\label{eq:nonaff2}
G_{A}-G_{NA}=\frac{2}{5\pi}\frac{\kappa}{R_{0}} \phi_{c}
e^{\alpha_{T}(T_{c}-T)}\sqrt{\phi_{c}[e^{\alpha_{T}(T_{c}-T)}-1]}.
\end{equation}
According to this equation for the shear modulus $G(T)$, nonaffinity alone
(induced by disorder) causes the melting at a critical point $T_{c}$ with the
scaling
$\sim\sqrt{T_{c}-T}$, even without the effects of thermal
vibrations on the rigidity. Including the effect of thermal phonons in the same
way as was done in Eq.(\ref{eq:aff}), the full expression for $G(T)$ becomes:
\begin{equation}\label{eq:nonaff}
G=\frac{2}{5\pi}  (\frac{\kappa}{R_{0}} \phi_{c}
e^{\alpha_{T}(T_{c}-T)}\sqrt{\phi_{c}[e^{\alpha_{T}(T_{c}-T)}-1]}-
\frac{kT}{R_{0}^{3}}e^{-\alpha_{T}T}).
\end{equation}
The square-root cusp singularity $G\sim \sqrt{T_{c}-T}$ predicted by our theory for the shear modulus has been
reported in the most recent numerical simulations for the melting of colloidal glasses~\cite{wittmer} which thus confirm the
validity of our approach.
To assess the interplay and relative magnitude of nonaffinity and thermal
phonons, it is important to have an expression for the critical point
$T_c$ in terms of parameters
used in this analysis. Since $T_c$ is, in effect, the glass transition
temperature, the task of finding it explicitly for a general system remains
challenging. However, since our comparison will be with a particular
experimental
system of polymeric glass, we can in fact offer an expression in such a
case. For polymer chains of $n$ units, the average connectivity due to
intra-chain
covalent bonds is close to 2: $z_{\rm co} = 2(1-1/n)$ ({the total coordination
number is
$z=z_{\rm co} + z_{\rm LJ}$). The LJ are central forces, but the
covalent
bonds also put a constraint on the bond angle. The classical
Phillips-Thorpe analysis of marginal stability~\cite{phillips,thorpe} gives
the fraction of floppy modes $f=N_{\mathrm{floppy}}/3N$ in a purely covalent
network: $f=1- \textstyle{\frac{1}{3}} \left(  \textstyle{\frac{1}{2}} z_{\rm
co} +[2z_{\rm co}-3] \right)$,
where every $z_{\rm co}$-coordinated monomer contributes $2z_{\rm co}-3$
bending constraints,
in addition to $\frac{1}{2} z_{\rm co}$ stretching constraints.
In our case, upon adding the LJ inter-chain bonds into the counting,
the fraction of floppy modes becomes:
\begin{equation}
f=1- \textstyle{\frac{1}{3}} \left(  \textstyle{\frac{1}{2}} z_{\rm co} +
[2z_{\rm co}-3] + \textstyle{\frac{1}{2}} z_{\rm LJ}   \right).
\end{equation}
Keeping $z_{\rm co}$ fixed, since it is $T$-independent and fixed by the
polymer chemistry,
we set $f=0$ in the above equation and solve for the critical value of $z_{\rm
LJ}$
at the rigidity transition, obtaining: $z_{\rm LJ}^* = 12-5z_{\rm co}$.
Upon applying $z_{c}=z_{\rm co} + z_{\rm LJ}^*$, we obtain the
critical value of the total connectivity $z$ at which
the rigidity is lost: $z_c = 12 - 4z_{\rm co} = 12-8(1-1/n)$. } In other
words, in order for the amorphous polymer assembly to become solid (glass)
there need to be at least $z_{\rm LJ}^* = z_c-z_{\rm co} = 12-5z_{\rm co}$
LJ inter-chain bonds per monomer, in addition to the chain
connectivity. Now we convert $z_c$ into the
critical volume fraction
$\phi_c$, via ${\phi _c} = \phi _c^* - \Lambda \cdot {z_{\rm co}}$, where
$\phi
_c^*$ is the packing fraction in the limit $z_{\rm co}=0$. If the attraction is
weak or absent, as in a system of hard spheres, then $\phi _c^*\simeq 0.64$ as
for random packings.

Finally, using the expression for $\phi_c (\alpha_T T)$, and for $z_{\rm
co}(n)$, we obtain the glass transition temperature for chains
with degree of polymerization $n$:
\begin{equation}\label{eq:Tg}
{T_c} = \frac{1}{\alpha _T}(1 - C - \phi _c^* + 2\Lambda)
- \frac{{2\Lambda}}{\alpha _T n}.
\end{equation}
The first term is what remains for very long chains ($n \gg 2\Lambda/\alpha
_T$). The
expression (\ref{eq:Tg}) provides a theoretical foundation for the empirical
dependence of the glass transition on $n$, first discussed by
Flory~\cite{flory,sperling}. {For common polymers,
the experimental values of the factor $2\Lambda/{\alpha _T}$
~\cite{flory,sperling} are of order $10^{3}K$, and hence
$\Lambda \simeq0.1$. If we take $T_{c}\simeq 383$K, as for polystyrene
glass used in the experimental comparison below, this gives a reasonable value
of
$C \simeq 0.48$.}
This value implies that $\phi \simeq 0.61\cdot e^{-\alpha _T T}$, in dense
amorphous polystyrene.

\begin{figure}
\includegraphics[width=.8\linewidth]{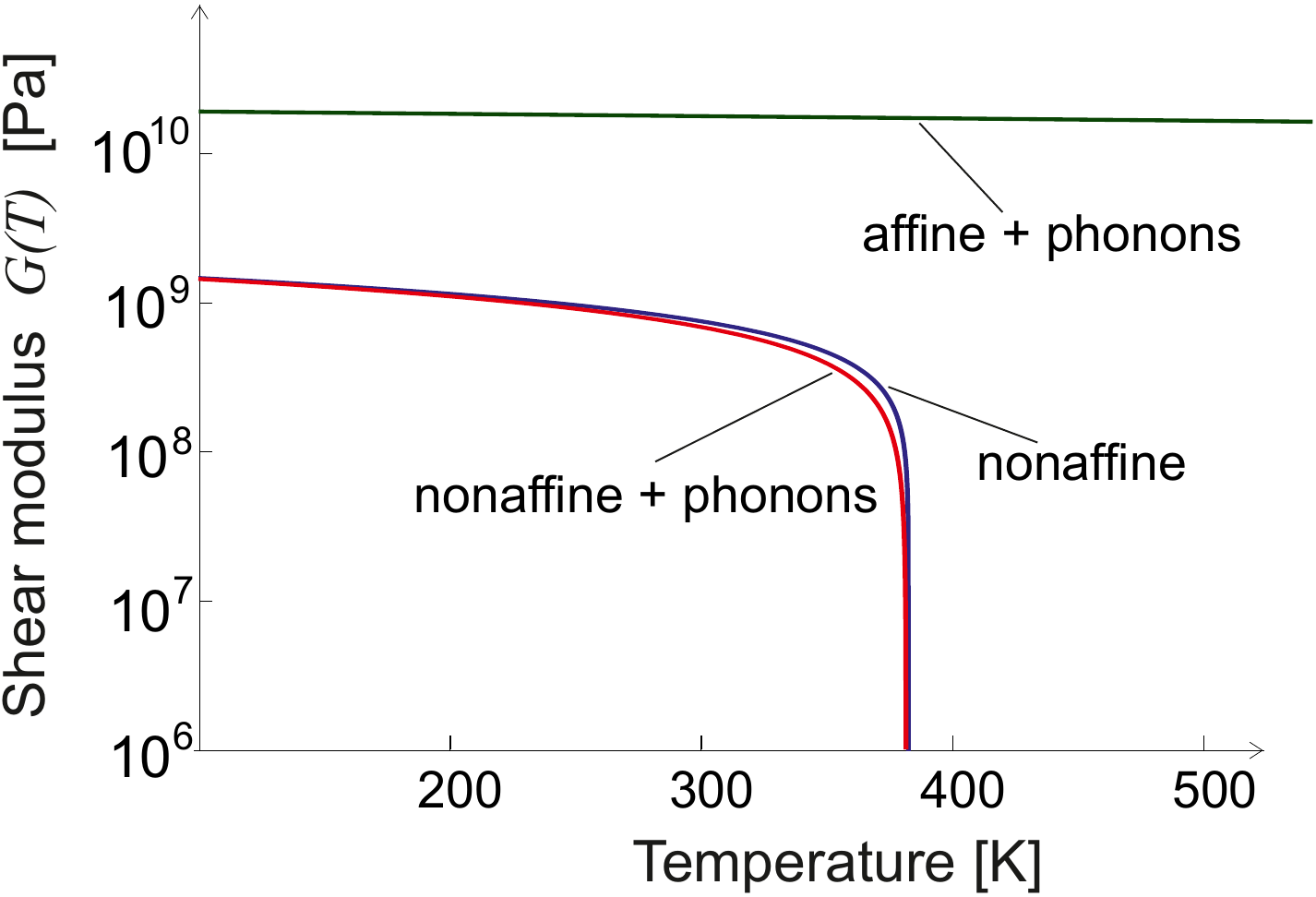}
\caption{\label{figure3}Comparison of theoretical predictions from
Eq.(\ref{eq:nonaff})
(nonaffine + phonons), Eq. (\ref{eq:nonaff2}) (nonaffine athermal), and Eq.
(\ref{eq:aff}) (affine + phonons). We take $T_{c}=383$K, $R_{0}=0.3$~nm,
$\alpha_{T}=2\cdot 10^{-4}\hbox{K}^{-1}$, as for polystyrene, ${\phi _c} = \phi
_c^* - \Lambda \cdot {z_{\rm co}}=0.44$, and the spring constant
$\kappa=50$~N/m.}
\end{figure}

\begin{figure}
\includegraphics[width=.85\linewidth]{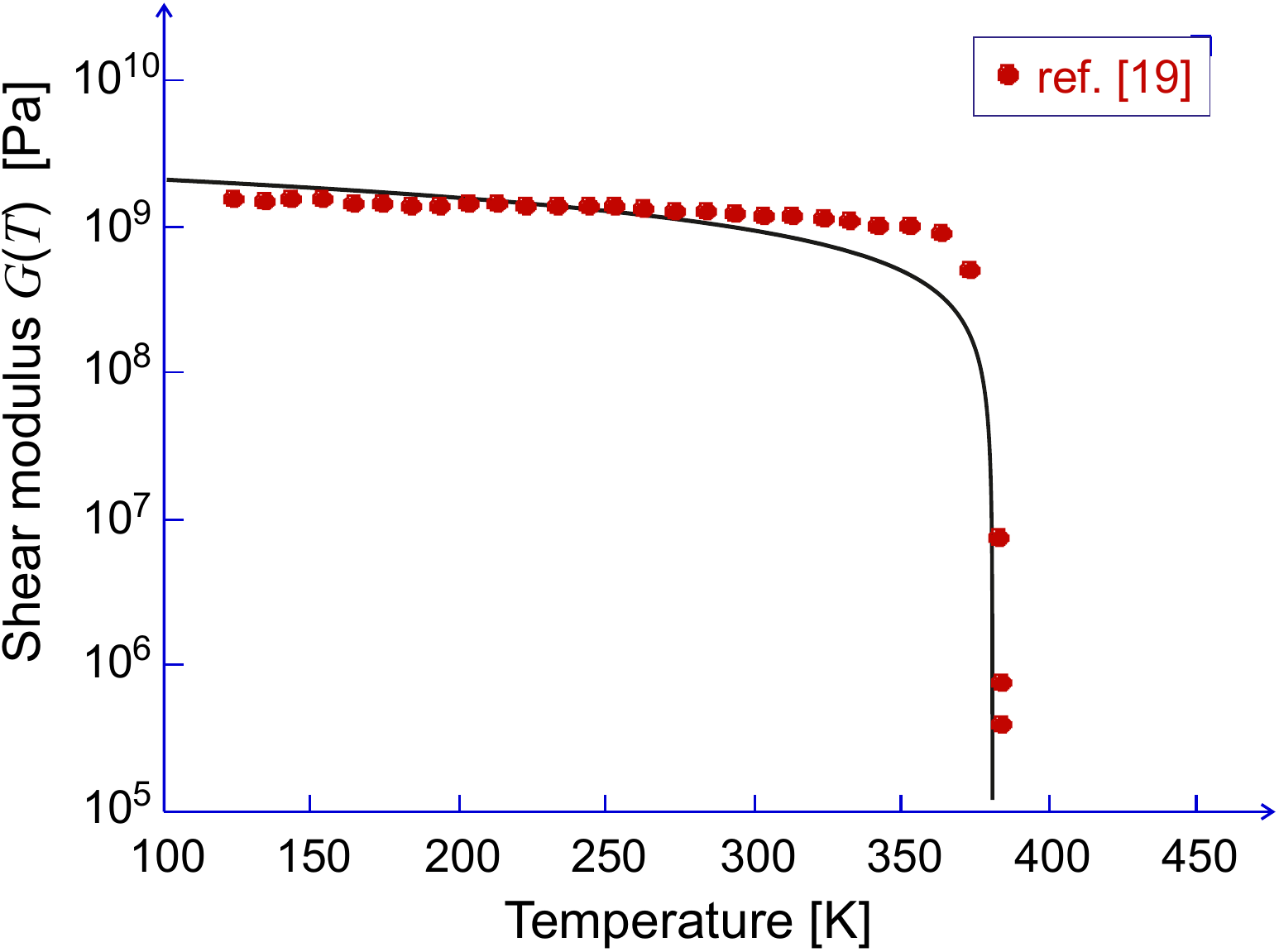}
\caption{\label{figure4}Theoretical fit of the experimental data for amorphous
polystyrene using Eq.(\ref{eq:nonaff}) and the same values for
parameters as in Fig.\ref{figure3}, including $T_{c}=383$K, and fitting
for the single parameter: the spring constant $\kappa=52$\,N/m. {The
experimental data were taken at a constant frequency of 0.9 Hz.}}
\end{figure}

In Fig.~\ref{figure3}  we have plotted predictions of different theories for
the
shear modulus $G(T)$ using the parameters of amorphous polystyrene, taking
$n=200$~\cite{schmieder}. It is evident that nonaffinity is the main
effect driving the melting transition, whereas thermal phonons have practically
no effect on the qualitative behavior of the melting
curve. On the other hand, if nonaffinity is neglected, the transition is
shifted to unrealistic,
enormously high temperatures.

Finally, Fig.~\ref{figure4} shows the comparison of our theory prediction for
the
melting of amorphous polystyrene with the classical experimental data of
Schmieder and Wolf~\cite{schmieder}, {reported also in Ferry's monograph as
representative of the quasistatic, low-frequency $G(T)$ in the glassy solid
state~\cite{ferry}.
As discussed by Ferry, the catastrophic drop of $G$ at the critical point is
the hallmark of the low-frequency (static) response, whereas at higher
frequency the drop becomes more gradual (due to the reduced nonaffinity, from
our point of view). Experimental data taken at even lower frequency would not
differ much as the drop can hardly be sharper than in Fig. 4.}

The value of the fitting parameter, the spring constant
$\kappa \approx 52$\,N/m is very sensible:
it corresponds to the C-C covalent bond enthalpy $350$kJ/mol over a distance
of $0.15$\,nm, which is almost exactly the C-C bond length ($0.146$\,nm).
More importantly, the theory can reproduce the qualitative behavior of the
experimental curve and the criticality (that are found also with many other
polymers~\cite{schmieder,ferry}) very well indeed. It is important to
notice
that no alternative
theory is available in the  literature for the mechanical response of amorphous
polymers at $T<T_{g}$ (in contrast, at  $T>T_{g}$, the reptation theory of de
Gennes, Doi and Edwards~\cite{doi} provides a good understanding of
viscoelasticity of
polymer melts).

From the point of view of applications, our theory can be used
to reconstruct the thermoelastic behavior of glassy
polymers at $T<T_{g}$, something which has not been possible so far, in the
same way as the Williams-Landel-Ferry theory~\cite{ferry} is used for the
viscosity at $T>T_{g}$.
The theory can be used to model also the melting curves of
inorganic materials, such as amorphous oxides, semiconductors and metallic
glasses where the thermal phonons play a more important role in view of the
much higher $T_{c}$.
%As a result, the theoretical framework
%presented here offers a completely new opportunity in material
%science: to quantitatively predict the thermo-mechanical response of
%materials and to control it by tuning the structural disorder
%(e.g. by means of structural defects).

\begin{acknowledgments}
We are grateful for discussions and input of
E. Scossa-Romano and F. Stellacci. This work has been supported by the
Ernest Oppenheimer Fellowship at Cambridge.
\end{acknowledgments}


\begin{thebibliography}{99}
\bibitem{gibbs} G. Adam and J.H. Gibbs, J. Chem. Phys. \textbf{43}, 139 (1965); E.A. DiMarzio and J.H. Gibbs, J. Polym. Sci. \textbf{40}, 121 (1959); E.A. DiMarzio and J.H. Gibbs, J. Polym. Sci. A \textbf{1}, 1417 (1963).
\bibitem{born} M. Born \emph{J. Chem. Phys.} \textbf{7}, 591-603 (1939).
\bibitem{edwards} S.F. Edwards and M. Warner \emph{Philos. Mag. A }\textbf{40},
257-278 (1979).
\bibitem{chaikin} P.M. Chaikin and T.C. Lubensky, \emph{Principles of Condensed
Matter Physics} (Cambridge University Press, Cambridge, 1995).
\bibitem{science} B.J. Siwick, J.R. Dwyer, R.E. Jordan, R.J.D.Miller
\emph{Science} \textbf{302}, 1382-1385 (2003).
\bibitem{frenkeld} S. Angioletti-Uberti, B. Mognetti, D. Frenkel \emph{Nature
Mater.}
\textbf{11}, 518-522 (2012).
\bibitem{rastogi} S. Rastogi, D.R. Lippits, G.W.M. Peters, R. Graf, Y.F. Yao,
H.W. Spiess \emph{Nature Mater}. \textbf{4}, 635-641 (2005).
\bibitem{brillouin} L. Brillouin \emph{Phys. Rev.} \textbf{54}, 916-917 (1938).
\bibitem{frenkel} J. Frenkel \emph{Kinetic Theory of Liquids} (Clarendon Press,
Oxford, 1946).
\bibitem{parodi} P.C. Martin, O. Parodi, and P.S. Pershan, Phys. Rev. A
\textbf{6}, 2401 (1972).
\bibitem{alexander} S. Alexander \emph{Phys. Rep.} {\bf 296}, 65-236 (1998).
\bibitem{lubensky} B.A. DiDonna, and T.C. Lubensky \emph{Phys. Rev. E} {\bf
72}, 066619 (2005).
\bibitem{barrat} A. Tanguy, J.P. Wittmer, F. Leonforte, and J.-L. Barrat
\emph{Phys. Rev. B} {\bf 66}, 174205 (2002).
\bibitem{mackintosh} D.A. Head, A.J. Levine, and F.C. MacKintosh \emph{Phys.
Rev. Lett.}
\textbf{91}, 108102 (2003).
\bibitem{mezard} H. Yoshino and M. Mezard \emph{Phys. Rev. Lett.} \textbf{105},
015504 (2010).
\bibitem{siebenbuerger} M. Siebenburger, M. Fuchs, H. Winter, and M. Ballauff,
J. Rheol. 53, 707 (2009);  G. Szamel and E. Flenner, Phys. Rev. Lett.
\textbf{107}, 105505 (2011).
\bibitem{barrat88} J.L. Barrat, J.N. Roux, J.P. Hansen, and M.L. Klein,
Europhys. Lett. \textbf{7}, 707 (1988).
\bibitem{klix} C.L. Klix, F. Ebert, F. Weysser, M. Fuchs, G. Maret, and P.
Keim, Phys. Rev. Lett. \textbf{109}, 178301.
\bibitem{thorpe1} M.F. Thorpe, J. Non-Crystalline Solids \textbf{57}, 355
(1983).
\bibitem{schmieder} K. Schmieder and K. Wolf \emph{Kolloid Z. Z.Polym.
}\textbf{134},
149-185 (1953).
\bibitem{zaccone} A. Zaccone \emph{J. Phys.: Condens. Matter}
\textbf{21}, 285103 (2009).
\bibitem{wyart} M. Wyart \emph{Ann. Phys. (Paris)} \textbf{30}, 1-96 (2005).
\bibitem{ohern} C.S. O'Hern, L.E. Silbert, A.J. Liu, and S. R. Nagel,
\emph{Phys. Rev. E }{\bf 68}, 011306
(2003).
\bibitem{zamponi} L. Berthier, H. Jaquin, and F. Zamponi, Phys. Rev. E 84,
051103 (2011).
\bibitem{lindemann} F.A. Lindemann \emph{Physik. Z.} \textbf{11}, 609-612
(1910).
\bibitem{lemaitre} A. Lemaitre  and  C. Maloney \emph{J.
Stat. Phys.} {\bf 123}, 415-453 (2006).
\bibitem{enzo} A. Zaccone and E. Scossa-Romano  \emph{Phys. Rev. B}
\textbf{83}, 184205 (2011)
\bibitem{jamie} A. Zaccone, J.R. Blundell, and E.M.Terentjev \emph{Phys. Rev.
B} \textbf{84},174119 (2011).
\bibitem{wittmer} J.P. Wittmer, H. Xu, P. Polinska, F. Weysser, J. Baschnagel, preprint at arXiv:1212.1593 [cond-mat.soft].
\bibitem{phillips} J.C. Phillips \emph{J. Non-Cryst. Solids}
\textbf{34}, 153-181 (1979).
\bibitem{thorpe}  H. He and M.F. Thorpe
\emph{Phys. Rev. Lett.} \textbf{54}, 2107-2110 (1985).
\bibitem{sperling} L.H. Sperling, Introduction to Physical Polymer Science,
(John Wiley and Sons, 2006).
\bibitem{flory} T.G. Fox and P.J. Flory J. Appl. Phys. \textbf{21}, 581-591
(1950).
\bibitem{ferry} J.D. Ferry,  \emph{Viscoelastic Properties of Polymers}, 3rd
ed. (Wiley and Sons, New York, 1980).
\bibitem{doi} M. Doi and S.F. Edwards,  \emph{The Theory of Polymer Dynamics}
(Clarendon Press, Oxford, 1986).




\end{thebibliography}
\end{document}